# Low-loss tunable infrared plasmons in the high-mobility perovskite (Ba,La)SnO$_3$

*Hongbin Yang,[1,*] Andrea Konečná,[2,3] Xianghan Xu,[4,5] Sang-Wook Cheong,[4,5] Eric Garfunkel,[1,4] F. Javier García de Abajo,[2,6] and Philip E. Batson[4]*

1. Department of Chemistry and Chemical Biology, Rutgers University, Piscataway, New Jersey, USA
2. ICFO-Institut de Ciencies Fotoniques, The Barcelona Institute of Science and Technology, 08860 Castelldefels, Barcelona, Spain
3. Central European Institute of Technology, Brno University of Technology, 61200 Brno, Czech Republic
4. Department of Physics and Astronomy, Rutgers University, Piscataway, New Jersey, USA
5. Rutgers Center for Emergent Materials, Rutgers University, Piscataway, New Jersey, USA
6. ICREA-Institució Catalana de Recerca i Estudis Avançats, Passeig Lluís Companys 23, 08010 Barcelona, Spain
Email: hongbin.yang@rutgers.edu

**Abstract**

BaSnO$_3$ exhibits the highest carrier mobility among perovskite oxides, making it ideal for oxide electronics. Collective charge carrier oscillations, plasmons, are expected to arise in this material, thus providing a tool to control the nanoscale optical field for optoelectronics applications. Here, we demonstrate the existence of relatively long-lived plasmons supported by high-mobility charge carriers in La-doped BaSnO$_3$ (BLSO). By exploiting the high spatial and energy resolution of electron energy-loss spectroscopy with a focused beam in a scanning transmission electron microscope, we systematically investigate the dispersion, confinement ratio, and damping of infrared localized surface plasmons (LSP) in BLSO nanoparticles. We find that the LSPs in BLSO are highly spatially confined compared to those sustained by noble metals and have relatively low loss and high quality factor compared to other doped oxides. Further analysis clarifies the relation between plasmon damping and carrier mobility in BLSO. Our results support the use of nanostructured degenerate semiconductors for plasmonic applications in the infrared region and establish a relevant alternative to more traditional plasmonic materials.

**Introduction**

Noble metals are the go-to choices for applications in plasmonics because of their relatively low optical losses and robustness[1-5], with intrinsic bulk plasmons emerging in the visible regime. Although surface plasmons in these materials can be pushed down to the near-infrared by shaping the materials into structures of high aspect ratio, this solution is not ideal, and alternative materials with their bulk plasmon frequency already in that spectral range would be useful. Additionally, traditional plasmonic metals do not allow for active tuning or ultrafast optical switching because of their high electron density, to which added doping charges can only contribute negligibly. Alkali metals present lower electron densities that place their plasmons in the near IR-visible region with extremely low losses[6], but these materials are unstable under ambient conditions and thus challenging to integrate in actual devices with long-term stability. The appeal of tunable carrier density, high carrier mobility, and good chemical stability has motivated the search for alternative plasmonic materials[7], including transparent conducting oxides (TCO)[8,9], transition metal nitrides[8,10], chalcogenides[11], and alloys[12], as well as two-dimensional materials[13], especially graphene[14-16] and black phorphous[17,18]. Among these, doped binary oxides, such as In$_2$O$_3$,

SnO$_2$, ZnO, and CdO, have been intensively studied in various geometries for their IR plasmonic properties[19-22]. More recently, semi-metallic perovskite oxides SrBO$_3$ (with B=Ge, V, or Nb)[23-26] have also been identified as alternative IR plasmonic materials.

The combination of appealing plasmonic and electronic properties in a single material adds extra versatility in the design of actual devices. In search of such matetrials, we consider BaSnO$_3$ (BSO), which is a wide band gap (2.9 to 3.2 eV)[27] perovskite oxide that holds great promises for applications in oxide electronics[28]. This material has a room-temperature carrier mobility >300 cm$^2$V$^{-1}$s$^{-1}$ in single crystals[29, 30], which is the highest amongst all transparent conductors and perovskite oxides, exceeding III-V semiconductors at high carrier densities[31]. This indium-free TCO also has exceptional stability of oxygen vacancies even under extreme biasing conditions[32] or at elevated temperatures[29]. Although several experimental works have studied the IR optical properties of bulk crystals[33], thin films,[34] and ensembles of nanocrystals[35], individual La-doped BSO (BLSO) nanoparticles with well-defined doping and shape have not been characterized yet. To that end, we exploit electron energy-loss spectroscopy (EELS) in a monochromated scanning transmission electron microscope (STEM), which allows for the spatial and spectral characterization of low-energy excitations at the single-particle level[36, 37].

In this work, we systematically study plasmons emerging in BLSO nanocrystals with well-defined shapes by measuring their spectral and spatial characteristics using state-of-the-art STEM-EELS[37, 38]. We observe infrared plasmons in the 50 - 800 meV energy range, and image their spatial distribution and localization in BLSO nanorods. We further explore the doping limit of La in BSO and the associated plasmon energies, which allow covering a wide range of IR frequencies reaching up to the telecom wavelength at 1.55 mm. In addition, we characterize the surface plasmon dispersion, confinement ratio, lifetime, and quality factor in individual nanoparticles with varying sizes. By comparing our results with recent studies of plasmons in conventional plasmonic metals (Au, Ag, Cu), we establish BLSO as an appealing plasmonic material more suitable to the infrared range, with some degree of tunability, and a low level of losses that correlates well with the carrier mobility.

**Results and discussion**

We synthetized nanocrystals of BLSO with varying sizes and geometries by a sol-gel method modified from Ref. [35] (Methods). Through atomic resolution high-angle annular dark-field (HAADF) in Fig. 1(a) and (b), we confirmed the expected cubic perovskite structure. We also observed that the BLSO particles have {100} terminated surfaces, the formation of which is likely related to the presence of fluorine on the surface. Core-loss EELS in Fig. 1(b) shows the presence of Ba, Sn, O, and La in the doped samples. In particular, La acts as n-type dopant that replaces the A-site Ba. We also investigated the doping limit of La in BSO and the maximum plasmon energy in this material with STEM-EELS. The bulk plasmon energy and the corresponding La percentage are extracted from low-loss and core-loss EELS, respectively. As shown in Fig. 1(c) and (d), both the bulk carrier plasmon energy and La $M_{4,5}$ edge intensity increase with nominal doping level. The low-loss and core-loss EELS signals were obtained from the same region, which allows correlating composition and plasmonic properties at the nanoscale. We find that highly doped BLSO can support plasmons up to an energy of 0.8 eV (corresponding to a wavelength of 1.55 µm), supporting BLSO as a promising candidate for plasmonics applications (such as epsilon-near-zero-based devices[39]) at the telecom wavelength with better optical properties than traditional TCOs because of the high carrier mobility of BLSO[34]. However, we find that doping inhomogeneity can play a role at high doping levels, and therefore, we limit our investigation to moderately (5%) La-doped BSO in the rest of this paper.

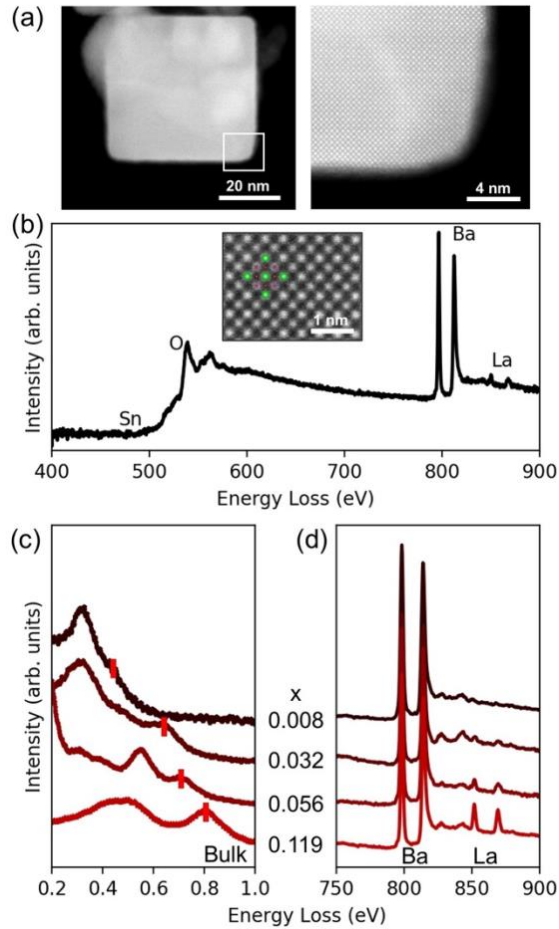

**Figure 1.** (a) Low magnification (left) and atomic resolution (right) HAADF-STEM image of a typical BLSO nanocrystal with rectangular cross section, taken along the [100] zone axis. (b) Core-loss EELS with edges indicating the presence of different elements. The inset shows a zoom-in view of the HAADF image in (a) with a structural model of $BaSnO_3$ superimposed on the image. (c, d) Evolution with doping level of the plasmon energy in the low-loss spectra and the intensity of the La $M_{4,5}$ edge in the core-loss spectra, respectively. Dopant percentages, x, determined from core-loss EELS, are shown as lables. Red markers in (c) denote bulk plasmons.

Besides the peaks assigned to bulk plasmon excitations denoted by red markers in Fig. 1(c), additional peaks appear at lower energies. These can also be observed in the aloof EEL spectra shown in Fig. 2 (a), where surface plasmon resonances of free charge carriers in BLSO are probed when the electron beam is place just outside a BLSO nanorod. We attribute these resonances to different localized surface plasmon (LSP) modes, which are spatially imaged in the energy-filtered maps shown in Fig. 2(a) and 2(b). Regions of high intensity in the maps are associated with an accumulation of induced electric field associated with the LSP excitations, as inferred from theoretical simulations. The distribution of these hotspots strongly depends on geometry. In particular, in the rectangular BLSO nanorod shown in Fig. 2(a) [length $L = 427$ nm, aspect ratio (AR) ~7], we observe multiple LSPs oscillating along the long axis, starting with a dipolar mode with wavelength $\lambda_{n=1} \approx 2L$, and two higher-order modes ($\lambda_{n=2} \approx L$ and $\lambda_{n=3} \approx 2L/3$). A transverse LSP mode is also observed at energies above 400 meV. In the BLSO cube shown in Fig. 2(b) ($L = 50$ nm, AR ~1), we observe plasmon resonances of shorter wavelengths characterized by field and charge accumulation at the corners, edges, and faces of the cube[40-43].

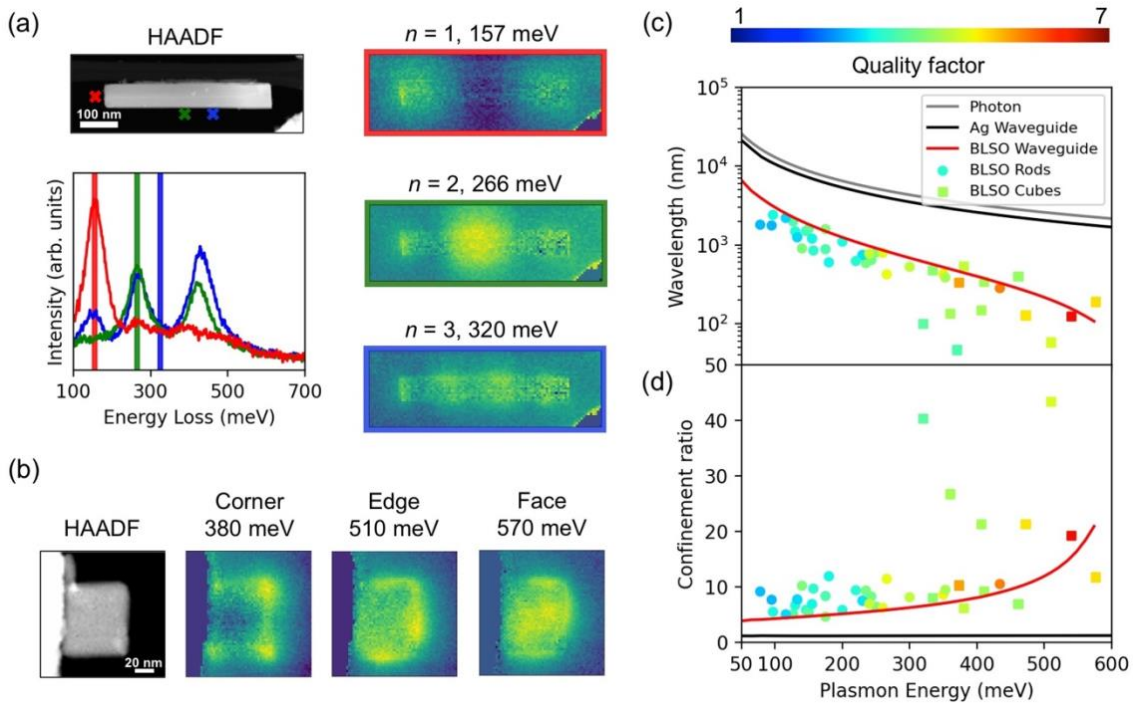

**Figure 2**. (a) Aloof EEL spectra and measured energy-filtered EELS maps for the infrared

localized surface plasmon modes in a BLSO nanorod. Electron beam positions are indicated by the color-matching crosses in the HAADF image. EELS maps are integrated in 10 meV wide windows marked in the EEL spectra. (b) EELS maps of LSPs in a BLSO nanocube for the labeled resonance energies. (c) Measured IR plasmon wavelengths in BLSO nanoparticles (symbols) compared with calculated values obtained for BLSO (red curve) and Ag (black curve) waveguides with a cross section of 50 × 50 nm$^2$. We also show the free-space photon dispersion (gray curve) for reference. (d) Confinement ratio obtained from (c) as the ratio of the plasmon excitation wavelength to the free-space photon wavelength.

One distinct difference between degenerate semiconductors and metals refers to the spatial confinement of LSPs with sub-eV energies. Noble metals present large ratios of the real and imaginary parts of the dielectric constant in the IR, which guarantees high quality factors of their LSP resonances. However, the corresponding plasmon wavelengths are very close to those of free-space photons in the IR because they host large densities of conduction electrons, so their bulk plasmons appear in the visible part of the spectrum, therefore demanding high ARs to move surface plasmons to the IR, where the dielectric function takes larger absolute values. In contrast, LSPs in doped semiconductors can produce a large confinement ratio in IR plasmons because of their low carrier densities, which lead to bulk plasmon energies already placed in the IR, so smaller ARs and lower values of the dielectric function already provide strongly confined plasmons in that region, in contrast to noble metals. With the plasmon energy and wavelength from the measured EELS energy-filtered maps in Fig. 2(c), we can readily obtain the plasmon confinement ratio in BLSO nanoparticles, as shown in Fig. 2(d), which turns out to be about one order of magnitude larger than for noble metals. In Fig. 2(c), we demonstrate that LSPs in BLSO nanoparticles have wavelengths ~6 to 12 times shorter than the photon wavelength at the same energy (but we note that a factor of 20 can be reached with the corner modes in nanocubes). The large spatial confinement is equally evident from the plasmon dispersions of BLSO calculated for an infinite waveguide with a square cross section (see also Fig. S2 in SI for calculations with different cross sections), which deviates away from the light line much more than the dispersion calculated for Ag waveguides of similar characteristics.

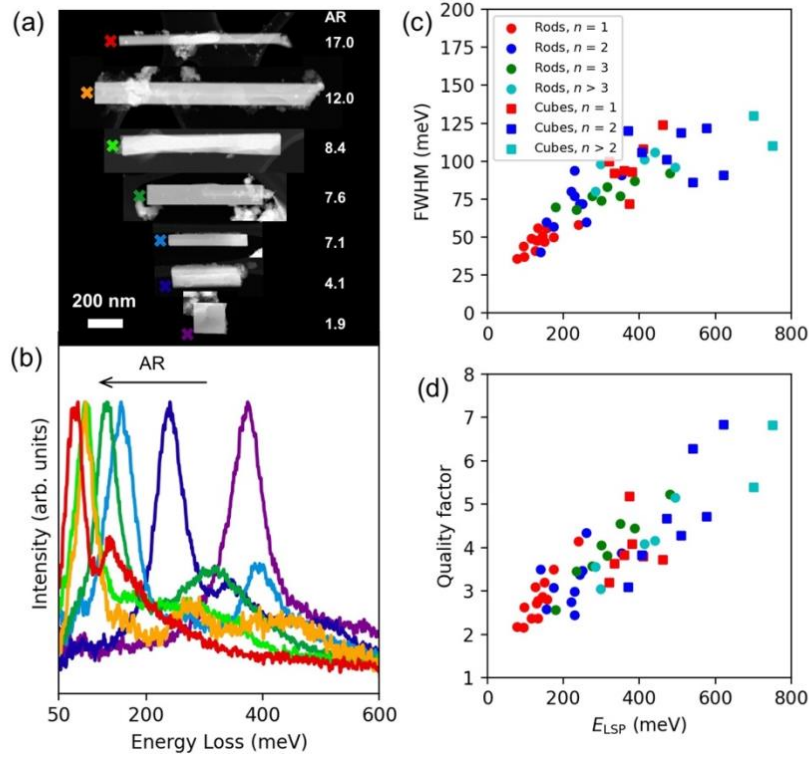

**Figure 3.** (a) HAADF-STEM images of BLSO rod-like particles with varying lengths and ARs (see labels). (b) Aloof EEL spectra acquired for beam positions as indicated by the color-matching crosses in (a), near the tip of the BLSO nanorods. (c) LSP FWHM and (d) quality factor as a function of plasmon energy.

Next, we study plasmon spectra for the series of BLSO nanorods shown in Fig. 3(a), which have a similar cross section and varying aspect ratio. These rod-like particles have rectangular cross sections and lengths of ~200 to 1200 nm, with AR ranging from 4 to 12. They are supported in part by lacey carbon TEM grids, while mostly suspended in vacuum. The EEL spectra shown in Fig. 3(b) were acquired with the beam placed just outside the nanorod tip or cube corner, as indicated by the colored markers in Fig. 3(a). With the help of the spatial mapping demonstrated in Fig. 2(a), we attribute the sharpest and most intense resonance peaks present in all spectra to the dipolar modes ($n = 1$). While the dipolar LSPs decrease in energy with increasing AR, their spectral width also decreases substantially. For a quantitative assessment of this trend, we performed Lorentzian fitting to extract the FWHM of all peaks corresponding to the observed LSPs, including the higher-order LSPs. As shown in Fig. 3(c), the FWHM can be as small as 35 meV for the nanorod with the largest AR and increases to about 120 meV for higher-energy LSPs. The quality factors Q (defined as the ratio of the plasmon-energy to the FHWM) obtained for these plasmons are shown in Fig. 3(d). The dipolar modes are found to have Q factors in the 2-5 range, while higher-energy LSPs show Q up to 8. Compared with other doped semiconductor nanocrystals[44-46], the LSPs observed in this study exhibit the largest Q factors in the mid-IR.

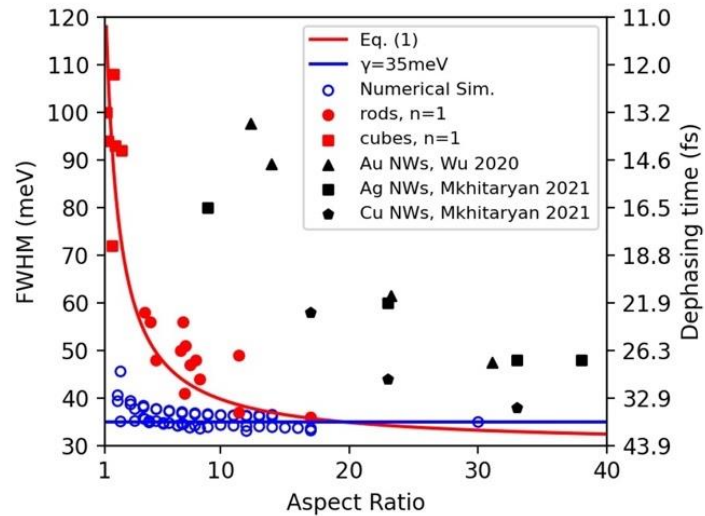

**Figure 4.** FWHM of the dipolar plasmon and the corresponding dephasing time as a function of aspect ratio. Experimental data from dipolar LSPs in BLSO nanorods and cubes (filled red circles and squares) are compared with numerical simulations for particles with similar lateral dimensions but varying cross sections (blue open circles). Results from literature for noble metal nanorods[47, 48] (black) are shown for comparison. The horizontal

blue line indicates the fitted Drude damping of our LBSO samples, $\hbar\gamma = 35$ meV.

While plasmon damping is often associated with carrier mobility of the material, such correlation is not generally direct, as several factors other than carrier mobility (e.g., nonlocal effects, surface quality, and radiative coupling) can play an essential role in determining the plasmon lifetime. We notice such variation of LSP FWHM with particle size and geometry in Fig. 3(c) and further study the plasmon FWHM as a function of nanoparticle AR and particle length $L$ in Fig. 4 and Fig. S4, respectively. We notice a substantial increase of LSP damping for BLSO particles with decreasing AR. To understand the possible origin of this trend, we calculate aloof EEL spectra for varying nanoparticle dimensions and extract the FWHM of the theoretically predicted peaks. We describe the optical properties of BLSO using a Drude dielectric function, which we find is sufficient for describing the charge carriers in doped BSO (Fig. S2), and perform numerical modelling as described in Methods. Also shown in Fig. 4, we find that the calculated FWHM of the dipolar plasmons increases only slightly above the bulk damping value $\hbar\gamma_{bulk} = 35$ meV here considered for the model, and in particular, the increase is ~5 meV for small AR or $L$. In the opposite limit, for very large AR or $L$, the FWHM of the dipolar plasmons only moves by 2-3 meV below the Drude damping. However, we stress that the calculated results deviate substantially from the experimental observations, especially when the particle sizes (and so the AR too) are small.

To further understand this discrepancy in the spectral widths, we calculated the optical scattering and absorption spectra near BLSO nanorods (SI) to analyze the origin of the plasmon damping. These quantities are roughly proportional to the cathodoluminescence (CL) and EELS probability, respectively, so we show in Fig. S5(a,b) the intensity ratio between these two probabilities (EELS/CL), which takes large values for the range of AR and $L$ studied here, suggesting a dominantly nonradiative character of the plasmonic excitations. We therefore conclude that damping in these excitations is dominated by material properties (i.e., intrinsic effects of the material and its modification due to the specific geometry of the particles), while radiative damping plays a negligible role.

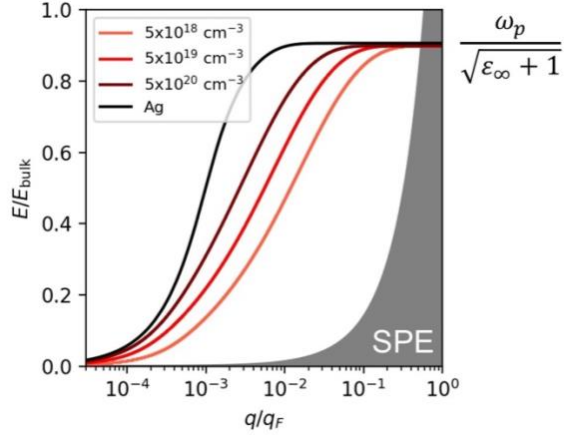

**Figure 5.** Normalized surface-plasmon dispersions of BLSO with carrier densities $N_{BLSO}=5\times10^{18}$, $5\times10^{19}$, and $5\times10^{20}$ cm$^{-3}$ (colored curves) and Ag (conduction electron density $N_{Ag}=6\times10^{22}$ cm$^{-3}$, black curve), as calculated for an infinite slab geometry. The vertical and horizontal axes are the normalized energy and the wave vector, where $E_{\text{bulk}}$ and $q_F$ are the bulk plasmon energy and Fermi wave vector, respectively. Single-particle excitation regime is indicated in gray.

To compare plasmon damping in BLSO with other plasmonic materials, it is illuminating to look at their respective normalized plasmon dispersions (Fig. 5). The surface plasmon dispersion relations are evaluated for infinite slabs with thickness of 50 nm[49]. We adopt the Drude model permittivity $\varepsilon(\omega) = \varepsilon_\infty - \omega_p^2/\omega(\omega + i\gamma_p)$ with parameters $\varepsilon_\infty = 4.0$ and hbar $E_{\text{bulk}} \equiv \hbar\omega_p/\sqrt{\varepsilon_\infty} = 4.6$ eV for silver, and $\varepsilon_\infty = 4.5$ and $\omega_p^2 = e^2 N/\varepsilon_0 m^*$ for BLSO, where $N$ is the doping carrier density (see legend in Fig. 5) and $m^*/m_e = 0.2$ is the carrier effective mass[50]. We plot the plasmon energy $E \equiv \hbar\omega$ as a function of wave vector $q$, normalized to $E_{\text{bulk}}$ and the Fermi wave vector $q_F$, respectively. For the same ratio $E/E_{\text{bulk}}$ of the plasmon energy $E$ to the bulk plasmon energy $E_{\text{bulk}}$, plasmons sustained by free carriers with lower density possess a higher ratio of the wave vector $q = 2\pi/\lambda$ to the Fermi wave vector $q_F$ compared to Ag, and this behavior extends up to the long-wave-vector limit, where the plasmon frequency converges to the non-retarded limit $\omega_p/\sqrt{\varepsilon_\infty + 1}$. In addition to their high spatial confinement, this indicates that BLSO plasmons are closer to the single-particle excitation (SPE) regime. It has been observed that surface plasmons exhibit increasing damping with increasing $q$, even before reaching the onset of SPEs at $q = q_F$[51]. For increasing wave vectors, shorter-wavelength plasmons are more likely to scatter off the doping inhomogeneities, neighboring nanoparticles, and shape irregularities. The differences in LSP energy between our simulations and experiments (Fig. S3) might be related to these extra scattering mechanisms, which are not contemplated in the theory. Furthermore, we investigate LSPs up to the high $E/E_{\text{bulk}}$ regime, where size effects are known to affect both the energy and width of LSPs[52]. A phenomenological model has often been used to take into account the size effects via the expression[53]

$$\gamma = \gamma_{\text{bulk}} + \frac{Av_F}{L}, \qquad (1)$$

where $v_F$ is the carrier Fermi velocity, $A$ is a phenomenological parameter that depends on particle morphology, surface details, and the surrounding medium, and $L$ is the particle size. We find that $\hbar\gamma_{\text{bulk}} = 30$ meV and $A$ in the range of 2 to 3 provides a reasonable agreement of $\gamma$ as a function of $L$ in comparison with experiment, as shown in Fig. 4. Although such size dependence has been observed for small metal particles[54], the physical explanation of the phenomenon is somewhat controversial[55], with two possible explanations for the increase in damping with decreasing particle size associated with either additional surface scattering determined by a size-independent electron mean free path[53] or the increasing role of the inhomogeneous electron density profile at the surface[56]. The range of particle size in which we observe an increased in plasmon FWHM is relatively large compared to that of metal particles (~10 nm or less), an effect that can be possibly related to the lower density of free carriers in BLSO.

Nevertheless, the dipolar plasmons in BLSO discussed in Figs. 3 and 4 are systematically displaying smaller values of the FWHM than those in noble metals of the same size[47, 48]. The reported BLSO plasmons reach their FWHM limit at an AR between 10 and 20, which is 2 to 3 times smaller than that for Ag, Au, or Cu. The BLSO particles required to support IR LSPs are much smaller than those made of noble metals, which makes radiative damping negligible in the former. In addition, both the intrinsic carrier mobility and the particle size affect the damping of plasmons in degenerate semiconductors, including BLSO. At large sizes, the FWHM of LSPs are primarily limited by the carrier mobility. In contrast, in smaller particles, additional contributions to plasmon damping are observed. The narrowest LSP that we observe in BLSO has a FWHM of 35 meV (i.e., a dephasing time of 38 fs). This amounts to a carrier mobility $\mu = 160$ cm$^2$/V$^{-1}$s$^{-1}$ which falls within the range of carrier mobilities previously determined for this material [57, 58]. Given that the carrier mobility in BaSnO$_3$ can be as high as 300 cm$^2$/V$^{-1}$s$^{-1}$, it is highly possible that better synthetic approaches and fabrication processes will lead to even better IR plasmons in doped BSO with more compact dimensions compared to noble metals.

**Conclusions**

In summary, we systematically identified and characterized infrared localized surface plasmons in individual nanocrystals of La-doped BaSnO$_3$ by STEM-EELS. Our results show that infrared plasmons sustained by BLSO are superior in spatial confinement ratio compared to those in noble metals. We also demonstrate that with high enough La doping, BSO can have a sufficiently large density of free carriers for its bulk plasma frequency to reach the telecommunication wavelength at 1.55 µm/0.8 eV. In this study, we analyze LSPs in BLSO nanorods with varying length and aspect ratio and find that their supported IR plasmons exhibit small losses, primarily limited by carrier mobility. Our results emphasize the strong potential of this high-mobility perovskite oxide for application in infrared plasmonic devices.

**Methods**

BLSO synthesis

We follow a recipe for BaSnO$_3$ nanoparticles synthesis via a sol-gel approach modified from Ref. [35]. Chlorine salts of Ba, La and Sn are weighted in desired molar ratio added in a mixture of deionized water and ethanol. Citric acid is then added into the solution. The solution is then kept at 80°C under stirring for 30 minutes to aid complete dissolution. Overdosed Ethylen glycol is then added, and we heat the solution to 100°C for 2 hours until the gel is formed. Annealing at 600°C followed by annealing at 1000°C results in phase pure cubic BLSO. Nanorod-like and cubic shaped BLSO particles can often be found in the product. When doping is successful and relatively uniform, the powder has a greenish and blue appearance, for small and large particle size, respectively.

Another synthetic route involves the formation of BaSn(OH)$_6$ as the middle product[59, 60] via participation at room temperature, followed by high-temperature annealing to form perovskite BaSnO$_3$. For this route, the solution containing salts of Ba, La, and Sn is kept at 80°C under vigorous stirring, while a NaOH solution is added dropwise until reaching a pH >7. During this process, Argon purging is also required to prevent the formation of oxides. At this point, white participation should already form, which is BaSn(OH)$_6$. Subsequent annealing above 600°C in Argon allows BSO nanorods to be formed with large length and aspect ratio. However, we find that it is difficult to incorporate La dopant into the A-site via this approach. Thus, this method is not suitable for producing plasmonic BLSO.

## STEM-EELS

A Nion UltraSTEM 100 scanning transmission electron microscope is used in this work. The microscope features an HERMES electron monochromator to improve the energy resolution. EEL spectra and energy-filtered maps are taken with energy resolution between 10-12 meV. The EELS detector dispersions are 0.9 and 1.3 meV/pixel for point EELS spectra acquisition and EELS mapping, respectively. Aloof EELS data are recorded with a detector dwell time of 256 ms. We use EELS spectra taken in vacuum far away from any specimen or grid to obtain the elastic background, which is removed from the spectra containing the inelastic signal[61, 62].

## Numerical Simulations

We perform the theoretical calculations using the software Comsol Multiphysics (RF module)[63, 64], where we implement a line current representing the electron beam and calculate the frequency- and spatially-dependent loss probability $\Gamma(\omega)$ as[65]

$$\Gamma(\mathbf{R}_b, \omega) = \frac{e}{\pi \hbar \omega} \int dz \, \text{Re}\left\{ E_z^{\text{ind}}(\mathbf{R}_b, z, \omega) e^{\frac{-i\omega z}{v}} \right\}.$$

Here, we assume electrons moving with constant velocity $v$, inducing an electric field $\mathbf{E}^{\text{ind}}$ along the electron trajectory. The integral is performed along the beam trajectory, which is taken to be parallel to the $z$ axis, intersecting the $xy$ plane at $\mathbf{R}_b$.

**Acknowledgement**

H.Y. and P.E.B. acknowledge the financial support of the US Department of Energy, Office of Science, Basic Energy Sciences under award number DE-SC0005132. X.X. and S.W.C. were supported by the center for Quantum Materials Synthesis (cQMS), funded by the Gordon and Betty Moore Foundation's EPiQS initiative through grant GBMF10104, and by Rutgers University. FJGA acknowledges support from the European Research Council (789104-eNANO) and the Spanish MINECO (PID2020-112625GB-I00 and CEX2019-000910-S).